\documentclass[12pt]{article}

\usepackage{cite}
\usepackage{epsfig}
\usepackage{amsmath}
\usepackage{pstricks}
\def\mathswitch#1{\relax\ifmmode#1\else$#1$\fi}
\def\mathswitchr#1{\relax\ifmmode{\mathrm{#1}}\else$\mathrm{#1}$\fi}

\newcommand{\tev}{\,\, \mathrm{TeV}}
\newcommand{\gev}{\,\, \mathrm{GeV}}

\newcommand{\SLASH}[2]{\makebox[#2ex][l]{$#1$}/}

\newcommand{\pslash}{\SLASH{p}{.2}}

\newcommand{\Eslash}{\SLASH{E}{.3}\,}
\newcommand{\mycaption}[1]{\caption{\sl #1}}

\begin{document}
\thispagestyle{empty}

\def\thefootnote{\fnsymbol{footnote}}

\begin{flushright}
%preprint numbers
\end{flushright}

\vspace{1cm}

\begin{center}

{\Large\sc {\bf General analysis of signals with two leptons and\\[.5ex]
 missing energy at the Large Hadron Collider}}
\\[3.5em]
{\large\sc
Chien-Yi~Chen$^1$, A.~Freitas$^2$
}

\vspace*{1cm}

{\sl $^1$ Department of Physics, Carnegie Mellon University, Pittsburgh, PA
15213, USA
\\[1em]
\sl $^2$
Department of Physics \& Astronomy, University of Pittsburgh,\\
3941 O'Hara St, Pittsburgh, PA 15260, USA
}

\end{center}

\vspace*{2.5cm}

\begin{abstract}

A signal of two leptons and missing energy is challenging to analyze at the
Large Hadron Collider (LHC) since it offers only few kinematical handles.
This signature generally arises from pair production of heavy charged particles
which each decay into a lepton and a weakly interacting stable particle.
Here this class of processes is analyzed with minimal model assumptions
by considering all possible combinations of spin 0, $\frac{1}{2}$ or 1, and of weak
iso-singlets, -doublets or -triplets for the new particles.
Adding to existing work on mass and spin measurements, two
new variables for spin determination and an asymmetry for the
determination of the couplings of the new particles are introduced.
It is shown that these observables allow
one to independently determine the spin and the couplings of the new particles,
except for a few cases that turn out to be indistinguishable at the LHC. These
findings are corroborated by results of an alternative analysis strategy based
on an automated likelihood test.

\end{abstract}

\setcounter{page}{0}
\setcounter{footnote}{0}

\newpage

%%%%%%%%%%%%%%%%%%%%%%%%%%%%%%%%%%%%%%%%%%%%%%%%%%%%%%%%%%%%%%%%%%%%%%%%%%%%%%

\section{Introduction}

Many models beyond the standard model (SM) include stable weekly interacting
massive particles which could be constituents of dark matter. Since the
stability of these particles is generally related to some symmetry, they can be
produced only in pairs at colliders, leading to challenging signatures with at
least two invisible objects. At hadron accelerators like the
Large Hadron Collider (LHC) such a signal is not sufficiently kinematically
constrained to enable the use of direct  reconstruction techniques, and thus it is very
difficult to uniquely determine the properties of the produced particles.

One of the most challenging cases are
processes with a low-multiplicity final state of only two
visible objects, which is focus of this article.
In particular we will consider the production of a pair of oppositely charged
heavy new particles $Y^\pm$ at the LHC, which each decay into a SM lepton and an
invisible neutral massive particle $X^0$:
\begin{equation}
pp \to Y^+Y^- \to \ell^+\ell^- X^0 \bar{X}^0, \qquad (\ell = e,\mu).
\label{eq:proc}
\end{equation}
Several methods for determining the $Y$ and $X$ masses in processes of this
type  have been proposed in the literature \cite{mt2,mass2,mt2isr,memnew}.
Furthermore, a number of authors have studied how to extract spin information
from angular distributions \cite{Barr:2005dz,maos,Buckley:2007th} and the total
production rate \cite{spinx}\footnote{The latter method, while potentially very
powerful, requires knowledge of the branching frac\-tions of $Y^\pm$, which are
\emph{a priori} unknown without model assumptions.}. To the best knowledge of
the authors, however, the problem of determining the couplings of the new
particles, which are related to their gauge group representations, has not yet
been considered.

The goal of this article is to analyze the process \eqref{eq:proc} in a more
model-independent approach by considering all possible assignments for spins (up
to spin one) and SU(2) representations (up to triplets) for the particles $X$
and $Y$. To discriminate between these template model combinations, we discuss
several variables for the measurement of masses, spins, and interactions
of the new particles,
including two new spin-sensitive observables and one new observable for the
coupling determination.
To minimize model dependence, the total cross section is not considered in
this set of variables.

Besides using dedicated observables, we also study an alternative analysis
strategy based on an automated likelihood test. This method matches the observed
lepton momenta in a sample of events to the corresponding momenta of a
theoretically calculated matrix element in a given model and calculates a
likelihood from that \cite{matrix,memtev}. Note that the two approaches based on
specific observables and on the automated likelihood test are complementary. The
latter often reaches a higher statistical significance due to the fact that no
information is lost by projecting onto some variables, but it does not permit a
straightforward separation between individual particle properties, such as spin
and couplings.

After characterizing all relevant spin and coupling representations in
section~\ref{setup}  and identifying 11 independent model combinations, we
present observables for the measurement of particle properties in
section~\ref{obs} and demonstrate their usefulness in a Monte-Carlo study.
Section~\ref{mem} is devoted to the analysis of the same set of template models
with the likelihood test method. Finally, our conclusions are given in
section~\ref{concl}.

%%%%%%%%%%%%%%%%%%%%%%%%%%%%%%%%%%%%%%%%%%%%%%%%%%%%%%%%%%%%%%%%%%%%%%%%%%%%%%

\section{Setup}
\label{setup}

\begin{figure}
\centering
\psfig{figure=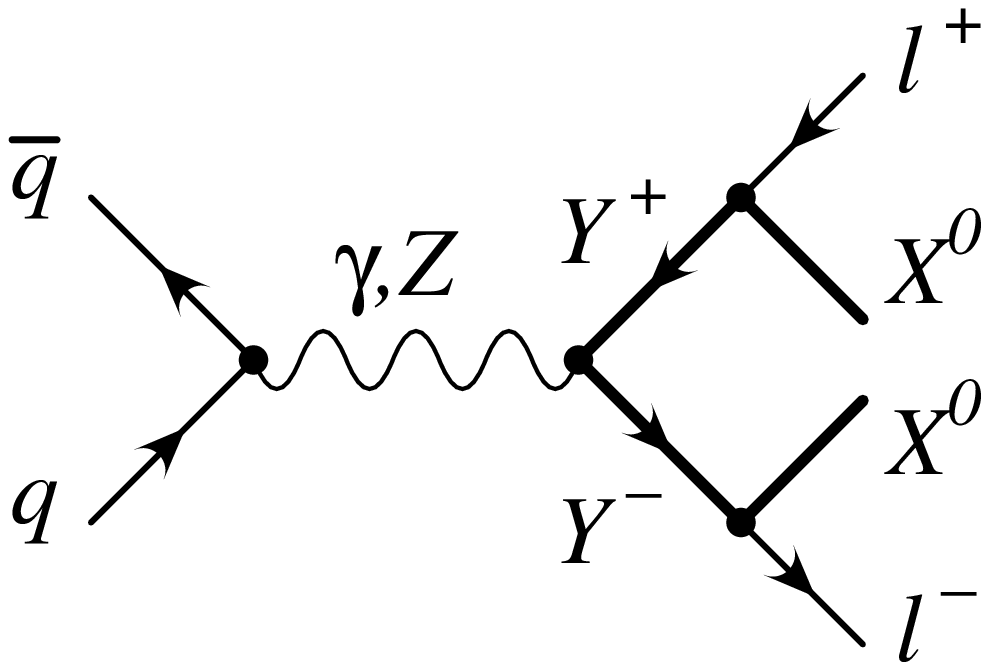, width=5.5cm}
\vspace{-.7em}
\mycaption{Basic diagram topology for the new physics processes under
consideration. Thick lines indicate new particles, while thin lines denote SM
particles.}
\label{fig:diag}
\end{figure}

The class of processes under consideration each involve Drell-Yan--type
production of a pair of charged heavy particles $Y^\pm$, which each subsequently
decay into a lepton $\ell^\pm$ ($\ell = e,\mu$) and a neutral heavy particle
$X^0$, see Fig.~\ref{fig:diag}. It is assumed that the $Y^\pm$ and $X^0$ are charged
under some discrete symmetry, such that they can be produced only in pairs and
the lighter new particle ($X^0$) is stable and escapes from the detector without
leaving a signal. The observable signature thus consists of two same-flavor
opposite-sign leptons and missing momentum: $\ell^+\ell^- + \Eslash$.
For this process it is insubstantial whether $X^0$ is self-conjugate or not.

%-----------------------------------------------------------------------------
%-----------------------------------------------------------------------------

For the purpose of this work
it is assumed that no other new heavy particles play a role in the s- or
t-channel of the $Y^+Y^-$ production process [note, however, that if $Y$ is a
vector boson this assumption is not valid, as will be explained below]. 
In fact, LHC data itself will be able to set a strong lower
bound on such particles: searches for di-jet resonances could rule out 
s-channel resonances that couple to light quarks up to several TeV, and any new
particles in the t-channel need to be colored and thus could be produced
directly with large cross sections unless their masses are larger than 2--3~TeV
\cite{lhcex}\footnote{These estimated bounds pertain to the LHC with a center-of-mass
energy of 14~TeV.}. Therefore, if the LHC does not see any such signals, one can
safely neglect the presence of extra particles in the s- and t-channel for the
production of $Y^+Y^-$ pairs with mass of a few hundred GeV.

%-----------------------------------------------------------------------------
\begin{table}[tp]
\renewcommand{\arraystretch}{1.4}
\centering
\begin{tabular}{|l||c|c|c||c|c||ll|}
\hline
 & $Y$ & $X$ & $\ell$ & $ZYY$ & $XY\ell$ & 
 \multicolumn{2}{l|}{sample model and decay} \\[-1ex]
 & $s$, $I_{\rm SU(2)}$ & $s$, $I_{\rm SU(2)}$ & $I_{\rm SU(2)}$ &
 coupling & coupling & \multicolumn{2}{l|}{$Y^- \to \ell^-X$} \\
\hline\hline
1 & 0, {\bf 1} & $\frac{1}{2}$, {\bf 1} & {\bf 1} & 
 $Z^\mu Y^* \!\overleftrightarrow{\partial}_{\!\!\!\mu} Y$ &
 $\overline{X}\frac{1+\gamma_5}{2}\ell \, Y^*$ &
 MSSM & $\tilde{\ell}_{\rm R}^- \to \ell^- \tilde{B}^0$ \\
\hline
1a & 0, {\bf 1} & $\frac{1}{2}$, {\bf 2} & {\bf 2} & 
 $Z^\mu Y^* \!\overleftrightarrow{\partial}_{\!\!\!\mu} Y$ &
 $\overline{X}\frac{1-\gamma_5}{2}\ell \, Y^*$ &
 MSSM & $\tilde{\ell}_{\rm R}^- \to \ell^- \tilde{H}^0$ \\
\hline
2 & 0, {\bf 2} & $\frac{1}{2}$, {\bf 1} & {\bf 2} & 
 $Z^\mu Y^* \!\overleftrightarrow{\partial}_{\!\!\!\mu} Y$ &
 $\overline{X}\frac{1-\gamma_5}{2}\ell \, Y^*$ &
 MSSM & $\tilde{\ell}_{\rm L}^- \to \ell^- \tilde{B}^0$ \\
\hline
2a & 0, {\bf 2} & $\frac{1}{2}$, {\bf 2} & {\bf 1} & 
 $Z^\mu Y^* \!\overleftrightarrow{\partial}_{\!\!\!\mu} Y$ &
 $\overline{X}\frac{1+\gamma_5}{2}\ell \, Y^*$ &
 MSSM & $\tilde{\ell}_{\rm L}^- \to \ell^- \tilde{H}^0$ \\
\hline
2b & 0, {\bf 2} & $\frac{1}{2}$, {\bf 3} & {\bf 2} & 
 $Z^\mu Y^* \!\overleftrightarrow{\partial}_{\!\!\!\mu} Y$ &
 $\overline{X}\frac{1-\gamma_5}{2}\ell \, Y^*$ &
 MSSM & $\tilde{\ell}_{\rm L}^- \to \ell^- \tilde{W}^0$ \\
\hline
3 & 0, {\bf 3} & $\frac{1}{2}$, {\bf 2} & {\bf 2} & 
 $Z^\mu Y^* \!\overleftrightarrow{\partial}_{\!\!\!\mu} Y$ &
 $\overline{X}\frac{1-\gamma_5}{2}\ell \, Y^*$ &
 UED6 & $W_{H,(1)}^- \to \ell^- \nu_{(1)}^{}$ \\
\hline
4 & $\frac{1}{2}$, {\bf 1} & 0, {\bf 1} & {\bf 1} &
 $\overline{Y} \SLASH{Z}{.4}\, Y$ &
 $\overline{Y}\frac{1+\gamma_5}{2}\ell \, X$ &
 UED6 & $\ell_{S,(1)}^- \to \ell^- B_{H,(1)}^0$ \\
\hline
5 & $\frac{1}{2}$, {\bf 1} & 0, {\bf 2} & {\bf 2} &
 $\overline{Y} \SLASH{Z}{.4}\, Y$ &
 $\overline{Y}\frac{1-\gamma_5}{2}\ell \, X$ &
 UED & $\ell_{S,(1)}^- \to \ell^- H_{(1)}^0$ \\
\hline
6 & $\frac{1}{2}$, {\bf 1} & 1, {\bf 1} & {\bf 1} &
 $\overline{Y} \SLASH{Z}{.4}\, Y$ &
 $\overline{Y}\SLASH{X}{.4}\, \frac{1+\gamma_5}{2}\ell$ &
 UED & $\ell_{S,(1)}^- \to \ell^- B_{\mu,(1)}^0$ \\
\hline
7 & $\frac{1}{2}$, {\bf 2} & 0, {\bf 1} & {\bf 2} &
 $\overline{Y} \SLASH{Z}{.4}\, Y$ &
 $\overline{Y}\frac{1-\gamma_5}{2}\ell \, X$ &
 UED6 & $\ell_{D,(1)}^- \to \ell^- B_{H,(1)}^0$ \\
\hline
7a & $\frac{1}{2}$, {\bf 2} & 0, {\bf 3} & {\bf 2} &
 $\overline{Y} \SLASH{Z}{.4}\, Y$ &
 $\overline{Y}\frac{1-\gamma_5}{2}\ell \, X$ &
 UED6 & $\ell_{D,(1)}^- \to \ell^- W_{H,(1)}^0$ \\
\hline
8 & $\frac{1}{2}$, {\bf 2} & 0, {\bf 2} & {\bf 1} &
 $\overline{Y} \SLASH{Z}{.4}\, Y$ &
 $\overline{Y}\frac{1+\gamma_5}{2}\ell \, X$ &
 MSSM & $\tilde{H}^- \to \ell^- \tilde{\nu}$ \\
\hline
9 & $\frac{1}{2}$, {\bf 2} & 1, {\bf 1} & {\bf 2} &
 $\overline{Y} \SLASH{Z}{.4}\, Y$ &
 $\overline{Y}\SLASH{X}{.4}\, \frac{1-\gamma_5}{2}\ell$ &
 UED & $\ell_{D,(1)}^- \to \ell^- B_{\mu,(1)}^0$ \\
\hline
9a & $\frac{1}{2}$, {\bf 2} & 1, {\bf 3} & {\bf 2} &
 $\overline{Y} \SLASH{Z}{.4}\, Y$ &
 $\overline{Y}\SLASH{X}{.4}\, \frac{1-\gamma_5}{2}\ell$ &
 UED & $\ell_{D,(1)}^- \to \ell^- W_{\mu,(1)}^0$ \\
\hline
10 & $\frac{1}{2}$, {\bf 3} & 0, {\bf 2} & {\bf 2} &
 $\overline{Y} \SLASH{Z}{.4}\, Y$ &
 $\overline{Y} \frac{1-\gamma_5}{2}\ell \, X$ &
 MSSM & $\tilde{W}^- \to \ell^- \tilde{\nu}$ \\
\hline
11 & 1, {\bf 3} & $\frac{1}{2}$, {\bf 2} & {\bf 2} &
 $S[Z,Y,Y^*]$ &
 $\overline{X} \SLASH{Y}{.2}\,^*\frac{1-\gamma_5}{2}\ell$ &
 UED & $W_{\mu,(1)}^- \to \ell^- \nu_{(1)}$ \\
\hline
\end{tabular}

\vspace{1ex}
\begin{tabular}{l}
$
A \!\overleftrightarrow{\partial}_{\!\!\!\mu} B \equiv
A (\partial_\mu B) - (\partial_\mu A) B,
$\\
$
S[Z,Y,Y^*] \equiv 
  Z_\mu \, Y^*_\nu \!\overleftrightarrow{\partial}^{\!\!\!\mu} Y^\nu +
  Y_\mu \, Z_\nu \!\overleftrightarrow{\partial}^{\!\!\!\mu} Y^{*\nu} +
  Y^*_\mu \, Y_\nu \!\overleftrightarrow{\partial}^{\!\!\!\mu} Z^\nu
$
\end{tabular}
\vspace{-1ex}
\mycaption{List of different assignments of spin $s$ and SU(2) representations
for the charged field $Y$ and the neutral field $X$. We define $Y^-/Y^+$ to be
the particle/anti-particle.
Also shown are the structure of the couplings to the $Z$ boson and to SM
leptons, as well as examples for realizations of these assignments 
in known models.
MSSM refers to the Minimal Supersymmetric Standard Model, UED to (at least) one
universal extra dimension, and UED to (at least) two
universal extra dimensions. $\tilde{\ell}_{\rm R}^-$, $\tilde{\ell}_{\rm L}^-$, 
$\tilde{\nu}$, $\tilde{B}^0$, $\tilde{W}^{0,\pm}$, and $\tilde{H}$ denote the
superpartners of the right-handed charged lepton, left-handed charged lepton,
neutrino, U(1) gauge field, SU(2) gauge fields, and Higgs boson, respectively.
$\ell^-_{S,(1)}$, $\ell^-_{D,(1)}$, $\nu_{(1)}$, $\tilde{B}^0_{\mu,(1)}$,
$\tilde{W}^{0,\pm}_{\mu,(1)}$, and $\tilde{H}_{(1)}$, respectively, are the
first-level KK-excitations of these fields. $B_{H,(1)}^0$ and $W_{H,(1)}^0$ are
scalars stemming from one of the extra components of the higher-dimensional
gauge fields in UED. More details of these models can be found in
Refs.~\cite{susy,ued}.}
\label{tab:models}
\end{table}
%-----------------------------------------------------------------------------

Table~\ref{tab:models} lists 16 possible combinations of spins up to spin one
and singlet, doublet, or (adjoint) triplet representations under the weak SU(2)
for the fields $X$ and $Y$. We do not consider complex SU(2) triplets, since
they contain doubly charged particles, which would lead to a clearly
distinguishable signature. 

Also shown are the structure of the couplings between
the $Y$ and the $Z$ boson and between $X$, $Y$, and SM charged leptons. The
$\gamma YY$ couplings has the same form as the $ZYY$ coupling. The coupling
constants for the $ZYY$ coupling are shown in Table~\ref{tab:rza}, given in
terms of the ratio with respect to the $\gamma YY$ coupling, $R_{\rm ZA} \equiv
g(ZYY)/e$. The
strength of the $XY\ell$ coupling depends on the detail of the given model, and
it is only relevant for the overall decay branchings, but not for the shapes of
distributions.
We neglect corrections from electroweak symmetry breaking to the masses and
interactions of $X$ and $Y$. As a result, if $Y$ is a spin-1/2 fermion it
couples to the $Z$ boson only through non-chiral vector couplings.

%-----------------------------------------------------------------------------
\begin{table}[tb]
\renewcommand{\arraystretch}{1.2}
\centering
\begin{tabular}{|c|c|}
\hline
$I_{\rm SU(2)}$  & $R_{\rm ZA}$ \\
\hline\hline 
{\bf 1} & $-\tan\theta_W \approx -0.548$ \\
\hline 
{\bf 2} & $\cot(2\theta_W) \approx 0.638$ \\
\hline 
{\bf 3} & $\cot\theta_W \approx 1.824$ \\
\hline
\end{tabular}
\mycaption{Ratio $R_{\rm ZA}$ of the $ZYY$ to $\gamma YY$ coupling strength
for different SU(2) representations of $Y$.}
\label{tab:rza}
\end{table}
%-----------------------------------------------------------------------------

For illustration, Tab.~\ref{tab:models} also gives examples for concrete realizations of
all 16 spin and gauge group assignments within the Minimal Supersymmetric
Standard Model (MSSM) or models with universal extra dimensions. However, many
of these combinations can also be realized in other models.

A comment is on order regarding the combination 11 in the table. Taking only the
s-channel diagrams in Fig.~\ref{fig:diag}, the cross section for spin-1 $Y$ pair
production grows unboundedly for increasing partonic center-of-mass energy. This
is a result of incomplete SU(2) gauge cancellations. Gauge invariance requires
the presence of an additional new particle in the t-channel, which interferes
negatively with the s-channel contribution and thus preserves perturbative
unitarity. In the case of universal extra dimensions this role is played by
the KK-quarks. Therefore, for model 11 we include a new colored fermion
$\hat{Q}$ that is charged under the same discrete symmetry as $X$ and $Y$.
The coupling strength of the $q\hat{Q}Y$ interaction is prescribed by gauge
invariance: $g(\bar{q}\hat{Q}Y) = g$. While for consistency it is necessary to
incorporate this particle in the cross-section calculation,
it is still possible that it is too massive to be seen directly at the LHC,
$i.\,e.$ $m_{\hat{Q}} > {\cal O}$(TeV).

The cross sections for the Drell-Yan--type process in Fig.~\ref{fig:diag} for
the different models in Tab.~\ref{tab:models} range from a few fb to several
hundred fb for a center-of-mass energy of $\sqrt{s}=14$~TeV and $X/Y$ masses of
a few hundred GeV, see appendix. Thus one can expect several 100--10,000 events being produced with a total luminosity of 100~fb$^{-1}$.
Note that in this paper, as mentioned in the introduction, the total event rate
will not be used to discriminate between models, since it would require
knowledge of the $Y^\pm$ branching fractions.

At the LHC it is not possible to determine the polarizations of the
final-state leptons and $X$ particles. As a result, several pairs of
combinations in Tab.~\ref{tab:models} are indistinguishable from each other, 
since after summing over the spins of the external legs  their squared matrix
elements are identical. Those sets of look-alikes are (1,~1a), (2,~2a,~2b), (7,~7a),
and (9,~9a).
This leaves a total of 11 potentially distinguishable combinations, which will be
explored in more detail in the following.

These 11 combinations have been implemented into {\sc CompHEP 4.5.1}
\cite{comphep} and representative samples with a few thousand parton-level
events have been generated for each of them. Since we will not consider the
total cross section as a discriminative quantity, the exact values for the
$XY\ell$ coupling strength and the widths of the $Y$ particles are irrelevant.
However, the $Y$ widths have been chosen small enough so that diagrams with
off-shell $Y$ particles can be safely neglected, $i.\,e.$ $\Gamma_Y/M_Y \ll
1\%$.

As a first step, 
initial-state radiation and detector acceptance effects have been ignored in the
following analysis, but we discuss these contributions in section~\ref{sim}.

%%%%%%%%%%%%%%%%%%%%%%%%%%%%%%%%%%%%%%%%%%%%%%%%%%%%%%%%%%%%%%%%%%%%%%%%%%%%%%

\section{Observables for Determination of Particle Properties}
\label{obs}

The experimental information  in the signature $\ell^+\ell^- + \Eslash$ consists
of the 3-momenta of the leptons $\ell^+$ and $\ell^-$, which can be parametrized
in terms of their transverse momentum $p_{\rm T}$, pseudorapidity $\eta$ and
azimuthal angle $\phi$. Since the system is invariant under overall azimuthal
rotations, one can construct five independent non-trivial observables from this
data. In this work we will focus on the following five quantities:
\begin{align} 
M_{\rm T2} &= \min_{{\bf p}_{\rm T,X_1}^{} + {\bf p}_{\rm T,X_2}^{} = 
\SLASH{\mbox{\scriptsize\bf p}}{.2}_{\rm T}}
 \left\{ \max \,\bigl( m_{\rm T}^{\ell^+,X_1}, m_{\rm T}^{\ell^-,X_2} \bigr) \right\},
 \label{eq:mt2} \\[1ex]
\cos\theta^*_{\ell\ell} &= \tanh \frac{|\eta_{\ell^+} - \eta_{\ell^-}|}{2}
 = \tanh \frac{\Delta\eta_{\ell\ell}}{2},  
 \label{eq:barr} \\[1ex]
%m_{\ell\ell}  \label{eq:minv} \\
M_{\rm eff} &= p_{\rm T,\ell^+} + p_{\rm T,\ell^-} + \pslash_{\rm T}, 
 \label{eq:meff} \\[1ex]
\Delta\phi_{\ell\ell} &= |(\phi_{\ell^+} - \phi_{\ell^-}) \mod 2\pi| \label{eq:dphi}, \\[1ex]
A_{\ell^+\ell^-} &= \frac{ N(E_{\ell^-} > E_{\ell^+}) 
 - N(E_{\ell^+} > E_{\ell^-}) }{ N(E_{\ell^-} > E_{\ell^+}) 
 + N(E_{\ell^+} > E_{\ell^-}) }.  \label{eq:A2}
\end{align}
Here
$$
\bigl( m_{\rm T}^{\ell^\pm,X_i} \bigr)^2 \equiv m_X^2 + 2\bigl(
p_{\rm T,\ell^\pm}\sqrt{m_X^2 + p_{\rm T,X_i}^2} - {\bf p}_{\rm T,\ell^\pm}
\cdot {\bf p}_{\rm T,X_i}^{}\bigr)
$$ 
is the transverse mass of the lepton
$\ell^\pm$, assumed to be massless, and one neutral heavy particle $X_i$, $i=1,2$.
Furthermore, $\theta^*_{\ell\ell}$ is the polar angle between one lepton and the beam
axis in a frame in which the pseudorapidities of the two leptons obey
$\eta^*_{\ell^+} = - \eta^*_{\ell^-}$, and $N(E_{\ell^-} > E_{\ell^+})$ denotes
the number of events for which $\ell^-$ has a larger energy than $\ell^+$.

This choice of observables is guided by their role 
in determining different particle properties.
The first observable in eq.~\eqref{eq:mt2} is useful for mass measurement,
eqs.~\eqref{eq:barr}--\eqref{eq:dphi} are sensitive to the spins of the new
particles, and eq.~\eqref{eq:A2} provides information about their couplings.

\subsection{Mass determination}

The variable $M_{\rm T2}$ has been proposed for the measurement of particle
masses in events with two or more invisible objects in the final state
\cite{mt2}. $M_{\rm T2}$ and similar variables have been studied extensively in
the literature \cite{mass2}, and it was shown that in favorable circumstances
one can use these variables to determine both the parent mass $m_Y$ as well as
the mass of the invisible child $m_X$, in particular by including information
about initial-state radiation \cite{mt2isr}. In this paper, therefore, mass determination
will not be discussed any further, and the reader is referred to
Refs.~\cite{mt2,mass2,mt2isr} for more details.

\subsection{Spin determination}

A useful observable for determining the spin $s_Y$ of the $Y$ particles is
$
\cos\theta^*_{\ell\ell} = \tanh (\Delta\eta_{\ell\ell}/2),
$
see eq.~\eqref{eq:barr},
which was introduced by Barr in Ref.~\cite{Barr:2005dz}. It is based on the
observation that the final state leptons $\ell^\pm$ tend to go in the same
direction as their parent particles $Y^\pm$, since on average the $Y^\pm$ are
produced with a sizable boost if $m_Y \ll \sqrt{s}$. As a result, the
distribution of the lepton polar angle $\theta^*_{\ell\ell}$, in the frame where
the pseudorapidities of the two leptons are equal in magnitude, is strongly
correlated to the production angle $\theta^*$ between one of the $Y$ and
the beam axis in the center-of-mass frame.

The $\theta^*$ distribution is closely connected to $s_Y$.
For the spin-0 and spin-1/2 cases one finds a characteristic difference which
is immediately visible in the formulas
\begin{align}
\text{scalar $Y$ (spin 0):} \qquad & \frac{d\sigma}{d\cos\theta^*} 
 \propto 1-\cos^2\theta^*, \\
\text{fermion $Y$ (spin $\tfrac{1}{2}$):} \qquad & \frac{d\sigma}{d\cos\theta^*} 
 \propto 2 + \beta^2_Y(\cos^2\theta^*-1),
\end{align}
where $\beta_Y$ is the velocity of the produced $Y$ particles. For spin-1 pair
production the situation is more complex since here one necessarily needs to
take into account a new particle $\hat{Q}$ in the t-channel. Depending on its
mass $m_{\hat{Q}}$, the observable $\theta^*_{\ell\ell}$ distribution can be
similar to the spin-0 case or to the spin-1/2 case, or different from both,
as can be seen from the
numerical results shown in section~\ref{num}.
Therefore, in general, the observable \eqref{eq:barr} alone does not unambiguously
distinguish spin-1 from spin-0 or spin-1/2.

One advantage of the definition \eqref{eq:barr} is that it is invariant under
longitudinal boost, $i.\,e.$ the value of $\tanh (\Delta\eta_{\ell\ell}/2)$ does
not depend on the momentum fractions carried by the quark and anti-quark in the
collision.

\vspace{\medskipamount}
Here we propose two other observables for the determination of the $Y$ spin: 
the effective mass $M_{\rm eff}$ and the difference between the azimuthal angles
of the leptons, $\Delta\phi_{\ell\ell}$,
see eqs.~\eqref{eq:meff} and \eqref{eq:dphi}.
The connection between these variables and $s_Y$ can be understood from
the threshold behavior of the partonic cross section $q\bar{q} \to Y^+Y^-$.
If $Y^+$ and $Y^-$ are scalars they are produced in a p-wave and 
the cross section behaves like $\sigma \sim \beta_Y^3$ near
threshold. For fermionic $Y$, instead, the cross section grows faster near
threshold, $\sigma \sim \beta_Y$. Therefore the cross section for fermionic $Y$
pair production reaches its maximum at lower values of the $Y^+Y^-$ invariant
mass, $m_{YY}$, than the cross section for scalar $Y$
pair production. The effective mass $M_{\rm eff}$ is strongly correlated to the 
$Y$-pair invariant mass, and thus the $M_{\rm eff}$ distribution will peak at
larger values for fermionic $Y$ than for scalar $Y$ (assuming that $m_Y$ is
equal in both cases and known from measuring the $M_{\rm T2}$ distribution).

\begin{figure}
\epsfig{figure=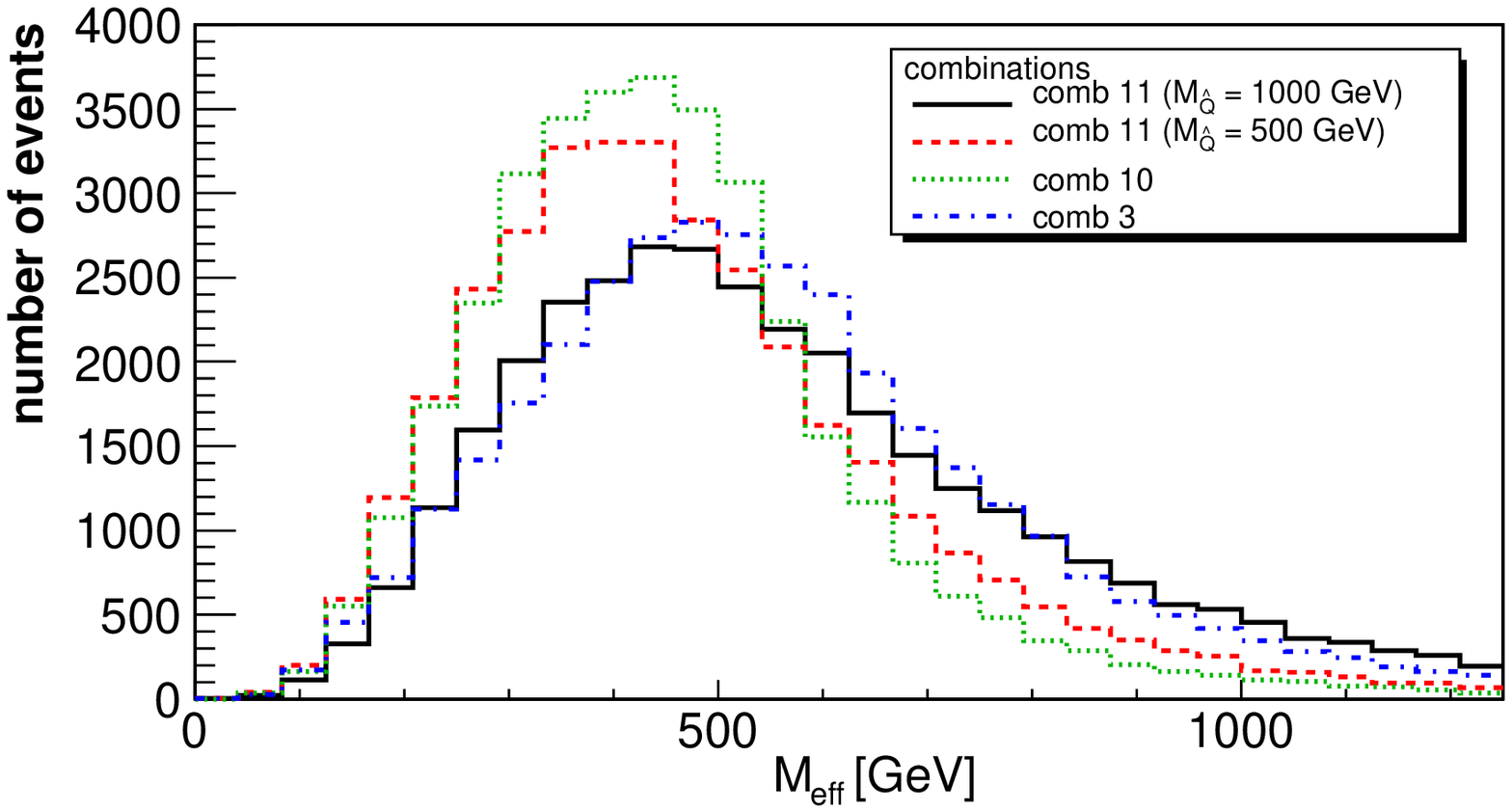, width=8cm, height=4.8cm, bb=55 10 539 277, clip=true}
\hfill
\epsfig{figure=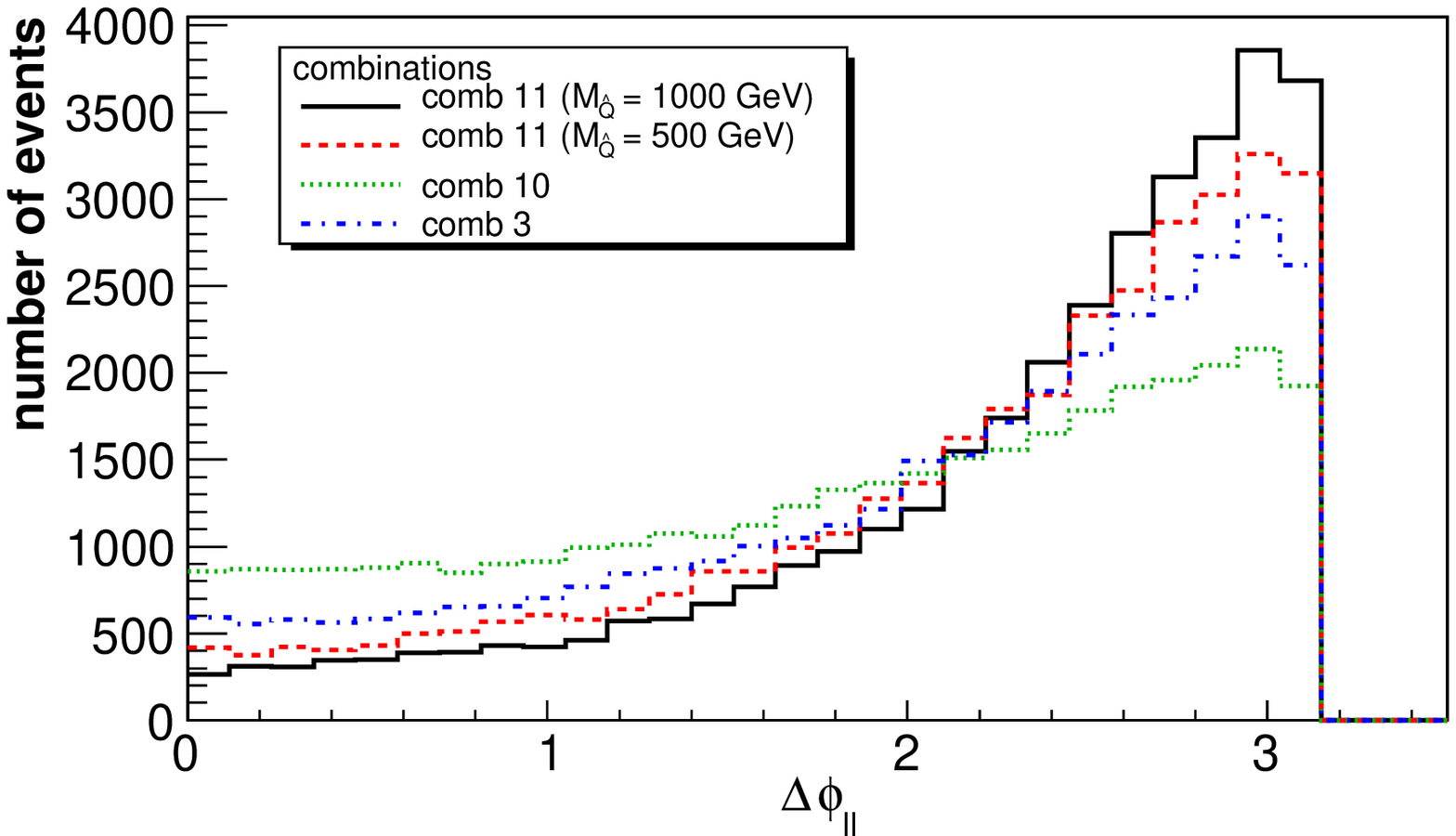, width=8cm, height=4.8cm, bb=55 10 539 300, clip=true}%
\vspace{-.3em}
\mycaption{Distributions for $M_{\rm eff}$ (left) and $\Delta\phi_{\ell\ell}$
(right), for combinations 3, 10, and 11 in Tab.~\ref{tab:models}, which correspond to $Y$
particles with spin 0, $\frac{1}{2}$, and 1, respectively. The plot is based on
35000 parton-level events for each combinations 
without cuts and detector effects, and using the mass
values $m_Y$~=~300~GeV and $m_X$~=~100~GeV. For combination 11, two choices of the
mass of the t-channel particles are shown, $m_{\hat{Q}}$~=~1000~GeV and 500~GeV.}
\label{fig:meffdphi}
\end{figure}

The dependence of the cross section on $m_{YY}$ also leaves a characteristic
imprint on the $\Delta\phi_{\ell\ell}$ distribution. Scalar $Y$ pairs will on
average be produced with a larger boost than fermionic $Y$ pairs.
This leads to a more pronounced peak at $\Delta\phi_{\ell\ell} \sim \pi$ in the
scalar case, since the larger boost is more likely to produce a back-to-back
configuration for the final-state leptons, see Fig.~\ref{fig:meffdphi}

%-----------------------------------------------------------------------------
%-----------------------------------------------------------------------------

If $Y^\pm$ are vector particles, the $M_{\rm eff}$ and $\Delta\phi_{\ell\ell}$
distributions depend on the mass $m_{\hat{Q}}$ of the particle in the t-channel.
For $m_{\hat{Q}} \gg m_Y$, the $Y^+Y^-$ pair production cross section reaches
its maximum at larger values of $m_{YY}$ than both the spin-0 and spin-1/2
cases, since the s-channel contribution alone grows monotonically with the
center-of-mass energy. In this case, the $M_{\rm eff}$ distribution for vector
$Y$ particles will peak at larger values than the other two cases, and the
$\Delta\phi_{\ell\ell}$ distribution will be very strongly peaked at $\pi$. On
the other hand, Fig.~\ref{fig:meffdphi} shows that for $m_{\hat{Q}}$ of the same
order as $m_Y$ the $M_{\rm eff}$ distribution can be similar to either  the
spin-0 or spin-1/2 cases, depending on the precise value of $m_{\hat{Q}}$.
Nevertheless, even for a relatively low value $m_{\hat{Q}} =
500\gev$\footnote{For such low values of $m_{\hat{Q}}$ one should see a signal
from direct production of the $\hat{Q}$ particle at the LHC.} the
$\Delta\phi_{\ell\ell}$ distribution is still distinctly different for spin-1
compared to the other spin cases. 
By using all three observables \eqref{eq:barr}--\eqref{eq:dphi} in combination one
therefore obtains the best discrimination power and can unambiguously
distinguish between $s_Y=0$, $\tfrac{1}{2}$, and 1.

Similar to $\tanh (\Delta\eta_{\ell\ell}/2)$, also $\Delta\phi_{\ell\ell}$ and
$M_{\rm eff}$  are invariant under longitudinal boosts, and thus very well
suited for hadron colliders.

\vspace{\medskipamount}
It needs to be pointed out that the three variables, $\tanh
(\Delta\eta_{\ell\ell}/2)$, $\Delta\phi_{\ell\ell}$ and $M_{\rm eff}$, are
primarily sensitive to the spin of the parent particle $Y$, but not of the
child particle $X$. Indeed, as can be seen from the numerical results in
sections~\ref{num} and \ref{mem}, it is very difficult to independently 
determine the $X$ spin.

\subsection{Coupling determination}

Experiments at LEP and SLC have determined the couplings of the $Z$ boson to SM
fermions with very high precision, in particular by measuring various left-right and forward-backward
asymmetries \cite{lep}.

Similarly, for the class of processes corresponding to Fig.~\ref{fig:diag},
one can in principle try to extract information about the $ZYY$ coupling by
constructing a forward-backward asymmetry for $pp \to Y^+Y^-$ at the
LHC. Although the initial $pp$ state is symmetric, the incoming quark for a
$q\bar{q}$-initiated process often stems from one of the valence quarks of
the protons and thus tends to have a larger momentum than the incoming
anti-quark. Therefore one can define the forward direction by the direction of
the overall longitudinal boost of an event.
However, since we neglect effects from electroweak symmetry breaking in the new
physics sector, all combinations in Tab.~\ref{tab:models} have parity-even
$ZYY$ couplings and the forward-backward asymmetry for $pp \to Y^+Y^-$ is
exactly zero.

On the other hand, the coupling between the incident $q\bar{q}$ pair and $Z$ boson has a
parity-odd axial-vector part, which results in the $Y^+Y^-$ pair being produced
with a non-vanishing \emph{polarization} asymmetry (unless $Y^\pm$ are scalars).
This polarization asymmetry can be probed through the decay $Y^\pm \to \ell^\pm
X^0$, since the interaction responsible for the decay is either left- or
right-handed and thus sensitive to the $Y$ polarization, see Tab.~\ref{tab:models}.

In the center-of-mass frame of the $Y^+Y^-$ system this leads to a
forward-backward asymmetry for the final-state leptons. As mentioned above, in
the lab frame the forward direction is defined by the overall boost of an event,
which is closely correlated to the direction of the more energetic of the two
leptons. Therefore we define the observable given in eq.~\eqref{eq:A2},
\begin{equation}
A_{\ell^+\ell^-} = \frac{ N(E_{\ell^-} > E_{\ell^+}) 
 - N(E_{\ell^+} > E_{\ell^-}) }{ N(E_{\ell^-} > E_{\ell^+}) 
 + N(E_{\ell^+} > E_{\ell^-}) }.  \tag{\ref{eq:A2}} \label{eq:A}
\end{equation}
The asymmetry is partially washed out by the mass $m_Y$, which can cause a
spin flip before the $Y$ decays, but we expect a non-vanishing result as long as
$m_Y \ll \sqrt{s}$.

%-----------------------------------------------------------------------------
\begin{figure}
\centering
\psfig{figure=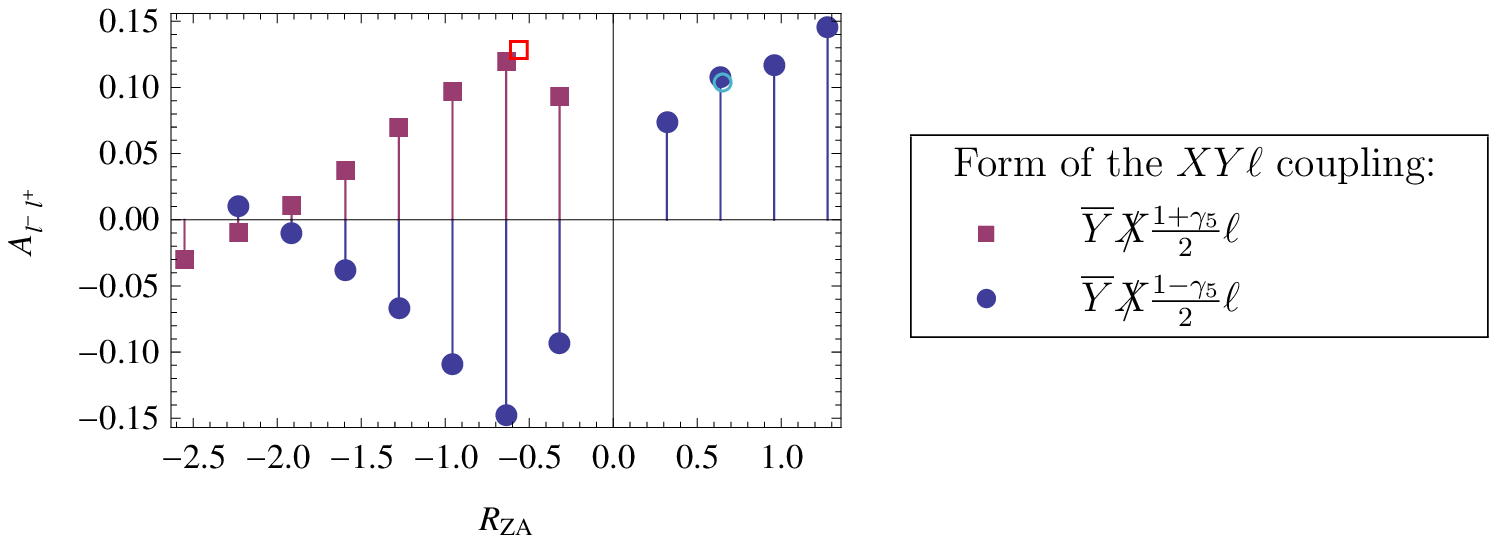, width=15cm, bb=30 525 460 680}
\vspace{-.7em}
\mycaption{Relationship between the coupling ratio $R_{\rm ZA} = g(ZYY)/e$ and
the asymmetry $A_{\ell^+\ell^-}$, for two different chiralities of the
interaction in the $Y^\pm \to \ell^\pm X^0$ decay, where $Y$ is a fermion and
$X$ is a vector boson. 
The combinations 6 and 9 in Tab.~\ref{tab:models} are indicated by the open square
and circle, respectively.
The plot is based on parton-level results without cuts and detector effects,
and using the mass values $m_Y$~=~300~GeV and $m_X$~=~100~GeV.}
\label{fig:asy}
\end{figure}
%-----------------------------------------------------------------------------

Eq.~\ref{eq:A} is mostly useful for discriminating between combinations with
$s_Y=\frac{1}{2}$, since scalars do not carry any polarization and lead to
a vanishing asymmetry, and there is only one combination with vector $Y$
particles in
Tab.~\ref{tab:models}. The value of $A_{\ell^+\ell^-}$ is connected to the size
of the ratio $R_{\rm ZA} = g(ZYY)/e$ between the $ZYY$ and $\gamma YY$ couplings
and to the sign of the $\gamma_5$ term in the $XY\ell$ coupling.
This is illustrated in Fig.~\ref{fig:asy} for the case that $Y$ is a fermion and
$X$ is a vector boson. As evident from the plot, there is a strong correlation
between $R_{\rm ZA}$ and $A_{\ell^+\ell^-}$, but one can encounter a two- to
three-fold ambiguity when trying to determine $R_{\rm ZA}$ from the measured 
value of $A_{\ell^+\ell^-}$.

\subsection{Numerical results}
\label{num}

Using {\sc CompHEP} we have generated parton-level events for all 11
combinations in Tab.~\ref{tab:models} for
a center-of-mass energy of $\sqrt{s}=14$~TeV and $m_Y = 300\gev$ and
$m_X=100\gev$. In the following discussion we will assume that the $Y$ and $X$
masses are known from observables like $M_{\rm T2}$, and for simplicity the uncertainty in the
mass determination will be neglected.

As explained in the previous section, the
observables $\tanh (\Delta\eta_{\ell\ell}/2)$, $M_{\rm eff}$, and
$\Delta\phi_{\ell\ell}$, see eqs.~\eqref{eq:barr}--\eqref{eq:dphi} can be used
to determine the spin $s_Y$ of the parent particle $Y$. We have checked this by
performing a 5-bin $\chi^2$ analysis for a sample of 5000 events for each
model combination, assuming Poisson statistics for the statistical error.
Table~\ref{tab:spin} shows the results for
combinations 3 (with
scalar $Y$), 10 (with fermion $Y$), and 11 (with vector $Y$) as examples.  
For the case of vector $Y$ particles, results for two sample values of the
t-channel fermion mass $m_{\hat{Q}}$ are given.
As evident from the table, by combining the three observables, one can
distinguish all three spin combinations from each other with a significance of
more than 18 standard deviations, for the given choice of masses and total event count.
This is true even for relatively small values of $m_{\hat{Q}} \sim {\cal O}(0.5
\tev)$.

We
have checked that the results are very similar if combinations 3 or 10 are
replaced by any of the other combinations with spin-0 or spin-1/2 $Y$ particles,
respectively. Furthermore, for any two models with identical $s_Y$ 
the distributions for all three variables are statistically consistent,
irrespective of the spin of $X$.

%-----------------------------------------------------------------------------
\begin{table}[tb]
\renewcommand{\arraystretch}{1.3}
\centering
\begin{tabular}{|l||c|c|c|c|c|}
\hline
 & \multicolumn{5}{c|}{(model A, model B)} \\
\cline{2-6}
Variable & (3,10) & 
  \parbox{1.01in}{\centering (3,11) [$M_{\hat{Q}}$=1~TeV]} & 
  \parbox{1.01in}{\centering (3,11) [$M_{\hat{Q}}$=0.5~TeV]} & 
  \parbox{1.01in}{\centering (10,11) [$M_{\hat{Q}}$=1~TeV]} & 
  \parbox{1.01in}{\centering (10,11) [$M_{\hat{Q}}$=0.5~TeV]} \\
\hline
$\tanh (\Delta\eta_{\ell\ell}/2)$ & 19.0 & 18.6 & 26.0 & 2.4 & 8.0 \\
$M_{\rm eff}$ & 37.5 & 3.9 & 25.1 & 30.7 & 9.5 \\
$\Delta\phi_{\ell\ell}$ & 16.3 & 21.4 & 10.7 & 41.1 & 29.0 \\
\hline
All combined & 37.5 & 18.6 & 26.0 & 41.1 & 29.0 \\
\hline
\end{tabular}
\mycaption{$\sqrt{\chi^2}$ values for a 5-bin $\chi^2$-test 
to discriminate between pairs of model
combinations with different spin of the parent $Y$ particle. The
combinations 3, 10, and 11 from Tab.~\ref{tab:models} have been chosen as examples
of models with $Y$ particles of spin 0, $\frac{1}{2}$, and 1, respectively.
{\rm Model~B} is assumed to represent the simulated ``data'', while {\rm
model~A} is the test hypothesis.
The results are based on samples of 5000 parton-level events without cuts and detector effects,
and using the input values $m_Y$~=~300~GeV, $m_X$~=~100~GeV, and
$\sqrt{s}$~=~14~TeV.}
\label{tab:spin}
\end{table}
%-----------------------------------------------------------------------------

To discriminate between models with identical $s_Y$ but different SU(2)
representations of the $Y$ and $X$ particles one can take advantage of the charge
asymmetry $A_{\ell^+\ell^-}$ in \eqref{eq:A2}.
As mentioned in the previous section, one cannot obtain a non-zero asymmetry if
the $Y$ particles are scalars, and we have confirmed this statement explicitly with our
simulation results.
However, for $s_Y=\frac{1}{2}$, $A_{\ell^+\ell^-}$ can yield useful
information about the structure of the $ZYY$ and $XY\ell$ couplings.
Results for the total asymmetry, without cuts or acceptance effects, are given
in Tab.~\ref{tab:asy} for all
combinations with fermionic $Y$ in Tab.~\ref{tab:models}.

%-----------------------------------------------------------------------------
\begin{table}[tb]
\centering
\begin{tabular}{|l|ccccccc|}
\hline
Combination from Tab.~\ref{tab:models} & 4 & 5 & 6 & 7 & 8 & 9 & 10 \\
\hline
$A_{\ell^+\ell^-}$ & $0.20$ & $-0.22$ & $0.13$ & $0.17$ & $-0.18$ & 
	$0.10$ & $0.20$ \\
\hline
\end{tabular}
\mycaption{Values for the asymmetry $A_{\ell^+\ell^-}$ for 
combinations with fermionic $Y$ in Tab.~\ref{tab:models}
based on simulated parton-level events for $m_Y$~=~300~GeV, $m_X$~=~100~GeV, and
$\sqrt{s}$~=~14~TeV.}
\label{tab:asy}
\end{table}
%-----------------------------------------------------------------------------

In general, $A_{\ell^+\ell^-}$ becomes maximal for events with large values of
$\cos\theta^*_{\ell\ell} = \tanh (\Delta\eta_{\ell\ell}/2)$ close to 1, $i.\,e.$
when the $Y^+Y^-$ pair is produced in the forward/backward direction. However,
this correlation between $A_{\ell^+\ell^-}$ and $\tanh
(\Delta\eta_{\ell\ell}/2)$ depends to a lesser extent also on spin effects in
the decay $Y \to \ell X$ and thus can be markedly different for models with
opposite chirality of the $XY\ell$ vertex. As a result, the significance for
distinction between such models  is increased by performing a binned analysis
for the distribution $dA_{\ell^+\ell^-}/d\tanh (\Delta\eta_{\ell\ell}/2)$. It
turns out that the highest sensitivity is obtained by using just two bins.

Table~\ref{tab:ferm} lists the statistical significances for discriminating
between any pair of the combinations 4--10 from Tab.~\ref{tab:models} based on this
observable. Models that have different signs for
$A_{\ell^+\ell^-}$ can be distinguished with more than 20 standard deviations
for an signal event sample of 5000 events (bold face numbers in the table).

However, the combinations 4, 7, and 10, as well as 6 and 9, 
are indistinguishable at the three-sigma level (gray italic numbers in the table).
It turns out that also when considering any other variables in
eqs.~\eqref{eq:mt2}--\eqref{eq:dphi}
one cannot achieve a higher significance for discriminating between these models.

%-----------------------------------------------------------------------------
\begin{table}[tb]
\centering
\renewcommand{\arraystretch}{1.3}
\begin{tabular}{|cc|cccccc|}
\cline{3-8}
\multicolumn{2}{c|}{} & \multicolumn{6}{c|}{model A} \\
\multicolumn{2}{c|}{} & 4 & 5 & 6 & 7 & 8 & 9 \\
\hline
 & 5 & {\bf 36} &&&&& \\
 & 6 & 4.9 & {\bf 29} &&&& \\
 & 7 & {\gray\sl 1.6} & {\bf 33} & 2.9 &&& \\
 & 8 & {\bf 32} & 3.5 & {\bf 26} & {\bf 29} &&\\
\rput[lt]{90}(0,0){\ \ model B}  & 9 & 6.7 & {\bf 27} & {\gray\sl 1.5} & 4.7 &
	{\bf 23} & \\
 & 10 & {\gray\sl 0.3} & {\bf 37} & 5.8 & {\gray\sl 2.2} & {\bf 33} & 7.6 \\
\hline
\end{tabular}
\mycaption{Statistical significance, in units of standard deviations,  for the
discrimination between combinations with fermionic $Y$ in Tab.~\ref{tab:models}
using the differential asymmetry $dA_{\ell^+\ell^-}/d\tanh
(\Delta\eta_{\ell\ell}/2)$. Numbers in bold face indicate a difference of
at least 20 standard deviations, while gray italic numbers 
denote a significance of less than three standard deviations. The results are
based on samples of 5000 parton-level events without cuts and detector effects,
and using the input values $m_Y$~=~300~GeV, $m_X$~=~100~GeV, and
$\sqrt{s}$~=~14~TeV.}
\label{tab:ferm}
\end{table}
%-----------------------------------------------------------------------------

Note that the variable $A_{\ell^+\ell^-}$ has some sensitivity to distinguish
between models which differ only through the spin of the $X$ particle, $i.\,e.$
between combinations 4 and 6, or 7 and 9 in Tab.~\ref{tab:models}. Assuming a
signal sample of 5000 events, as in Tab.~\ref{tab:ferm}, a discrimination
significance of about five standard deviations  can be achieved for these pairs.

\subsection{Simulation results}
\label{sim}

The analysis in the previous section does not take into account detector
acceptance and signal selection cuts. To study the influence of these effects
on the results we have passed the parton-level events generated by {\sc
CompHEP} \cite{comphep}  through {\sc Pythia 6.4} \cite{pythia} and {\sc PGS4}
\cite{pgs}. By including initial-state radiation and parton showering in the
{\sc Pythia} simulation one can furthermore evaluate whether fluctuations of the
initial-state transverse momentum might wash out the characteristic features for
the model discrimination.

In Ref.~\cite{Barr:2005dz} it has been shown that the selection cuts
\begin{align}
N(\ell^+) &= N(\ell^-) = 1, &
m_{\ell\ell} &> 150\gev, &
&\max\{p_{\rm T, \ell^\pm}\} > 40\gev, \quad \min\{p_{\rm T, \ell^\pm}\} > 30\gev, \nonumber \\
\pslash_{\rm T} &> 100\gev, &
M_{\rm T2} &> 100\gev, &
&|\SLASH{\mbox{\bf p}}{.2}_{\rm T} + {\bf p}_{\rm T, \ell^+}
 + {\bf p}_{\rm T, \ell^-}| < 100\gev,  \nonumber \\
p_{\rm T,j} &< 100\gev, &
N_b &= 0, \label{eq:cuts}
\end{align}
reduce the SM background rate to about 1.6~fb. Here $N(\ell^\pm)$ denotes the
number of visible leptons $\ell^\pm=e^\pm,\mu^\pm$ in the central detector,
$N_b$ denotes the number of vertex $b$ tags, $m_{\ell\ell}$ is the di-lepton
invariant mass, and $p_{\rm T,j}$ refers to the transverse momentum of any
reconstructed jet. With these cuts one obtains
a selection efficiency for the signal process
$pp \to Y^+Y^- \to \ell^+\ell^-XX$ between 27\% and 40\%, 
depending on the specific type of $Y$ and $X$ particle. As listed
in the appendix, this corresponds to measurable signal cross sections between
about 1~fb and 200~fb. For concreteness we will assume 5000 observed events,
which corresponds to the expected yield of model combination 7 for an integrated
luminosity of 200~fb$^{-1}$. In comparison, the SM background of about 300 events
is small and can be neglected.

%-----------------------------------------------------------------------------
\begin{table}[tb]
\renewcommand{\arraystretch}{1.3}
\centering
\begin{tabular}{|l||c|c|c|c|c|}
\hline
 & \multicolumn{5}{c|}{(model A, model B)} \\
\cline{2-6}
Variable & (3,10) & 
  \parbox{1.01in}{\centering (3,11) [$M_{\hat{Q}}$=1~TeV]} & 
  \parbox{1.01in}{\centering (3,11) [$M_{\hat{Q}}$=0.5~TeV]} & 
  \parbox{1.01in}{\centering (10,11) [$M_{\hat{Q}}$=1~TeV]} & 
  \parbox{1.01in}{\centering (10,11) [$M_{\hat{Q}}$=0.5~TeV]} \\
\hline
$\tanh (\Delta\eta_{\ell\ell}/2)$ & 23.2 & 20.1 & 30.6 & 6.3 & 10.1 \\
$M_{\rm eff}$ & 40.0 & 9.7 & 24.2 & 36.8 & 12.6 \\
$\Delta\phi_{\ell\ell}$ & 27.2 & 15.5 & 6.0 & 40.7 & 23.8 \\
\hline
All combined & 40.0 & 20.1 & 30.6 & 40.7 & 23.8 \\
\hline
\end{tabular}
\mycaption{$\sqrt{\chi^2}$ values for a 5-bin $\chi^2$-test 
to discriminate between pairs of model
combinations with different spin of the parent $Y$ particle, for a sample of
5000 events passing the detector simulation and selection cuts in
eq.~\eqref{eq:cuts}. The notation and input parameters are the same as in
Tab.~\ref{tab:spin}.}
\label{tab:spinc}
\end{table}
%-----------------------------------------------------------------------------
\begin{table}[tb]
\centering
\renewcommand{\arraystretch}{1.3}
\begin{tabular}{|cc|cccccc|}
\cline{3-8}
\multicolumn{2}{c|}{} & \multicolumn{6}{c|}{model A} \\
\multicolumn{2}{c|}{} & 4 & 5 & 6 & 7 & 8 & 9 \\
\hline
 & 5 & {\bf 23} &&&&& \\
 & 6 & 3.3 & {\bf 20} &&&& \\
 & 7 & {\gray\sl 2.0} & {\bf 22} & {\gray\sl 1.6} &&& \\
 & 8 & {\bf 22} & {\gray\sl 2.1} & 19 & {\bf 21} &&\\
\rput[lt]{90}(0,0){\ \ model B}  
 & 9 & 3.7 & 19 & {\gray\sl 1.7} & 3.3 & 17 & \\
 & 10& {\gray\sl 1.3} & {\bf 25} & 4.1 & {\gray\sl 2.1} & {\bf 23} & 5.3 \\
\hline
\end{tabular}
\mycaption{Statistical significance, in units of standard deviations,  for the
discrimination between combinations with fermionic $Y$
using $dA_{\ell^+\ell^-}/d\tanh (\Delta\eta_{\ell\ell}/2)$. 
The results are based on a sample of
5000 events passing the detector simulation and selection cuts in
eq.~\eqref{eq:cuts}, with notation and input parameters are the same as in
Tab.~\ref{tab:ferm}.}
\label{tab:fermc}
\end{table}
%-----------------------------------------------------------------------------

Tables~\ref{tab:spinc} and \ref{tab:fermc} summarize the significance for
distinguishing between models with different $Y$ spin and with different
couplings, assuming 5000 measured events for $\sqrt{s}=14$~TeV, $m_Y = 300\gev$
and $m_X=100\gev$. Overall, the obtained significances for the spin
discrimination are comparable to the parton-level results in
Tab.~\ref{tab:spin}, and in a few cases the significance is even higher. This
seemingly surprising outcome is related to the fact that we compare the same
number of ``observed'' events in the previous section and in this section, but in the latter
case the cuts remove part of the phase space, leaving a higher event yield in
the remaining phase-space region.

For the coupling determination one finds that the asymmetry $A_{\ell^+\ell^-}$
is washed out noticeably by the cuts, leading to substantially reduced
significances in Tab.~\ref{tab:fermc} compared to Tab.~\ref{tab:ferm}.
Nevertheless, models with different sign for $A_{\ell^+\ell^-}$ can still be
distinguished with at least 17 standard deviations.

In summary, for most cases, selection cuts and smearing effects only moderately
affect the capability for identifying particle properties with the
described observables. Of course the selection cuts reduce the overall event
number, which however also depends on the model-dependent total cross section
and thus is left as a free parameter here.

%%%%%%%%%%%%%%%%%%%%%%%%%%%%%%%%%%%%%%%%%%%%%%%%%%%%%%%%%%%%%%%%%%%%%%%%%%%%%%

\section{Comparison with Automated Likelihood Analysis}
\label{mem}

An alternative approach for the analysis of a new-physics signal 
is an automated likelihood test for a sample of measured
events.  With such a computerized procedure it is in general not possible
to clearly separate properties like spin and couplings, but it offers the
advantage of reaching a higher sensitivity by using the complete event
information instead of specific observables. A very appealing realization of an
automated likelihood analysis is the Matrix Element Method (MEM)
\cite{matrix,memtev},
which uses parton-level matrix elements to specify the theoretical model that is
compared with the data.  The method can be used to measure one or several parameters
of the model  by finding the maximum of the likelihood for a sample of events as
a function of these parameters.  As of today, the MEM achieves the most precise
determination of the top-quark mass \cite{memtev} and new-physics particle
masses \cite{memnew}.

For each single event, with observed momenta $\textbf{p}_i^{\rm vis}$, the MEM
defines a likelihood measure that it  agrees with a model for a given set of
model parameters $\alpha$:
\begin{align}
{\cal P}(\textbf{p}^{\rm vis}_i|\alpha) &= \frac{1}{\sigma_\alpha} \int dx_1 dx
_2 \,
 \frac{f_1(x_1)f_2(x_2)}{2sx_1x_2}
\left[ \prod_{i \in \text{final}} \int
 \frac{d^3 p_i}{(2\pi)^3 2E_i} \right] |M_\alpha(p_i)|^2
 \prod_{i \in \text{vis}} \delta(\textbf{p}_i - \textbf{p}^{\rm vis}_i).
 \label{eq:mem1}
\end{align}
Here $f_1$ and $f_2$ are the parton distribution functions, $M_\alpha$ is the
theoretical matrix element, and $\sigma_\alpha$ is the total cross section,
computed with the same matrix element.  The 3-momenta $\textbf{p}_i^{\rm
vis}$ of the visible measured objects are matched with the corresponding momenta
$\textbf{p}_i$ of the final state particles in the matrix element, while the
momenta of invisible particles (weakly interacting particles, such as the $X$
particle in our case) are integrated over.

For a sample of $N$ events, the combined likelihood is usually stated in terms
of its logarithm, which in the large-$N$ limit can be interpreted as a $\chi^2$
value,
\begin{equation}
\chi^2 = 
-2\ln({\cal L}) = -2\sum_{n=1}^N \ln {\cal P}(\textbf{p}^{\rm vis}_{n,i}|\alpha)
,
\label{eq:logl}
\end{equation}
where $\textbf{p}^{\rm vis}_{n,i}$ are the measured momenta of the $n$th
event.

The MEM is particularly useful for signals that cannot be fully reconstructed
due to invisible final-state particles,
and it can be applied to determine the masses of both $X$ and $Y$ in processes of
the type in eq.~\eqref{eq:proc} \cite{memnew}. Here we will assume that
the masses are already known and instead focus on the discrimination between the
models in Tab.~\ref{tab:models}.

Matrix elements for all 11 combinations in the table have been computed with the
help of {\sc CompHEP} and implemented into a private code for performing the
phase-space integration in \eqref{eq:mem1}. Similar to section~\ref{num} only
parton-level events without cuts have been used in this analysis.
Results for model comparisons are
listed in Tables~\ref{tab:spin2} and \ref{tab:ferm2}.

%-----------------------------------------------------------------------------
\begin{table}[tb]
\renewcommand{\arraystretch}{1.3}
\centering
\begin{tabular}{|c|c|c|c|c|}
\hline
 \multicolumn{5}{|c|}{(model A, model B)} \\
\hline
 (3,10) & 
  \parbox{1.01in}{\centering (3,11) [$M_{\hat{Q}}$=1~TeV]} & 
  \parbox{1.01in}{\centering (3,11) [$M_{\hat{Q}}$=0.5~TeV]} & 
  \parbox{1.01in}{\centering (10,11) [$M_{\hat{Q}}$=1~TeV]} & 
  \parbox{1.01in}{\centering (10,11) [$M_{\hat{Q}}$=0.5~TeV]} \\
\hline
 60 & 59 & 61 & 85 & 87 \\
\hline
\end{tabular}
\mycaption{Statistical significance, in units of standard deviations,  
for the discrimination between pairs of model
combinations with different spin of the parent $Y$ particle, based on the MEM. 
A sample of 5000 parton-level events without cuts and detector effects has been
used. 
The notation and input parameters are the same as in Tab.~\ref{tab:spin}.}
\label{tab:spin2}
\end{table}
%-----------------------------------------------------------------------------

As can be seen from Tab.~\ref{tab:spin2}, the MEM achieves a much higher
significance for discriminating between combinations with different $s_Y$, see
Tab.~\ref{tab:spin} for comparison. This is not surprising since several
observables, eqs.~\eqref{eq:barr}--\eqref{eq:dphi}, were found to be sensitive
to the $Y$ spin, indicating that none of them captures all relevant information.
Note also that the results in Tab.~\ref{tab:spin2} do not depend strongly on the
unknown mass of the t-channel fermion $\hat{Q}$ for combination~11.

%-----------------------------------------------------------------------------
\begin{table}[tb]
\centering
\renewcommand{\arraystretch}{1.3}
\begin{tabular}{|cc|cccccc|}
\cline{3-8}
\multicolumn{2}{c|}{} & \multicolumn{6}{c|}{model A} \\
\multicolumn{2}{c|}{} & 4 & 5 & 6 & 7 & 8 & 9 \\
\hline
 & 5 & {\bf 30} &&&&& \\
 & 6 & 8.3 & {\bf 25} &&&& \\
 & 7 & {\gray\sl 2.5} & {\bf 29} & 9.1 &&& \\
 & 8 & {\bf 30} & {\gray\sl 2.3} & {\bf 27} & {\bf 28} &&\\
\rput[lt]{90}(0,0){\ \ model B}  
 & 9 & 8.6 & {\bf 26} & {\gray\sl 3.0} & 9.0 &	{\bf 27} & \\
 & 10 & 15 & {\bf 41} & {\bf 22} & 14 & {\bf 43} & 20 \\
\hline
\end{tabular}
\mycaption{Statistical significance, in units of standard deviations,  for the
discrimination between combinations with fermionic $Y$ in Tab.~\ref{tab:models}
based on the MEM. A sample of 5000 parton-level events without cuts and detector effects has been
used. The notation and input parameters are the same as in Tab.~\ref{tab:ferm}.}
\label{tab:ferm2}
\end{table}
%-----------------------------------------------------------------------------

The MEM can also distinguish between combinations that all have spin-1/2 $Y$
particles but which differ in the SU(2) representations of $X$ and $Y$, as shown
in Tab.~\ref{tab:ferm2}. It is interesting to note that in most cases the
statistical significance achieved by the MEM is comparable to the results
obtained with the asymmetry $A_{\ell^+\ell^-}$ in Tab.~\ref{tab:ferm}. An
exception is combination 10 which can be distinguished  from the other
combinations with substantially higher significance using the MEM compared to
$A_{\ell^+\ell^-}$. This implies that the asymmetry $A_{\ell^+\ell^-}$ captures
essentially all measurable information about the $ZYY$ and $XY\ell$ couplings,
except for the special case of model 10.

Similar to the results of the previous section, it is found that one cannot
discriminate very well between combinations with $Y$ singlets and $Y$ doublets,
$i.\,e.$ between 4 and 7, 5 and 8, or 6 and 9\footnote{Note, however, that a better
differentiation between these cases would in principle be possible with more
statistics, requiring significantly larger amounts of integrated luminosity.}.
Likewise, the MEM results for the
combinations 1, 2, and 3 with scalar $Y$ differ by less than one standard
deviation, and thus are completely indistinguishable.

%%%%%%%%%%%%%%%%%%%%%%%%%%%%%%%%%%%%%%%%%%%%%%%%%%%%%%%%%%%%%%%%%%%%%%%%%%%%%%

\section{Conclusions}
\label{concl}

This paper presents a comprehensive analysis of
new physics processes of the form $pp \to Y^+Y^- \to \ell^+\ell^- X^0
\bar{X}^0\, (\ell = e,\mu)$, where $X^0$ is stable and weakly interacting,
leading to a signature of two opposite-sign same-flavor leptons and missing
momentum. To minimize model assumptions, all possible combinations for the spins
and weak SU(2) couplings of $X$ and $Y$ have been considered, allowing for spin 0,
$\frac{1}{2}$ and 1, and SU(2) iso-singlets, -doublets and -triplets, see
Tab.~\ref{tab:models}.

The signal processes have been analyzed with two different and complementary approaches.
The first method is based on specific observables. Concretely, we have studied three variables
for the measurement of the spins and one asymmetry for the extraction of
information about the couplings of the new particles.
Secondly, an automated strategy called the Matrix Element Method has been used,
which algorithmically computes a likelihood that a given event sample agrees
with some model interpretation supplied in the form of a theoretically
calculated matrix element.

It has been found that the spin $s_Y$ of the parent particle $Y$ can be
determined with high statistical significance, so that a sample of a few hundred
signal events is sufficient for discrimination at the 5$\sigma$-level.
Furthermore, it was shown that the asymmetry $A_{\ell^+\ell^-}$ defined in
eq.~\eqref{eq:A} is instrumental in distinguishing between model combinations
that all have $s_Y=\frac{1}{2}$ but different $Y$ and $X$ couplings. The
majority of possible coupling assignments can be differentiated with high
significance, but it turns out that for the pairs 4 and 7, 5 and 8, as well as 6 and 9 in
Tab.~\ref{tab:models} one cannot achieve a
$3\sigma$ discrimination with a realistic number of a few thousand events. This
is related to the fact that the relationship between the $ZYY$ coupling strength
and the observable asymmetry is not monotonic and can involve degenerate
solutions. Remarkably, the same model combinations are also difficult to
distinguish with the Matrix Element Method, which demonstrates that the
asymmetry $A_{\ell^+\ell^-}$ reflects all relevant information about the
couplings of the underlying model.

For $s_Y=0$ it is generally impossible to discriminate between cases with
different couplings or with different spin of the $X$ particles, due to the
absence of spin correlations between the production and decay stages of the
process. For $s_Y=1$ the coupling structure of the process is essentially fixed
by gauge invariance and thus already uniquely known once the vector nature of
$Y$ has been determined.

Our findings indicate that even for the challenging case of a process with a
short one-step decay chain it is in general possible to separately determine
the spins and couplings of the new heavy particles. The results in this paper
have been presented for the specific choice of masses $m_Y=300\gev$ and
$m_X=100\gev$, but we have checked explicitly that the essential features are
unchanged for $m_Y=200\gev$. While the main goal of this study was the
development of the theoretical framework and conceptual ideas, we have also
performed a fast detector simulation with selection cuts for the suppression of
standard model backgrounds and found that qualitatively our conclusions still
hold.
Nevertheless, a dedicated experimental simulation with a careful evaluation of
systematic errors, including the influence of uncertainties in the $Y$ and $X$
masses, would be required to check the
viability of our results under realistic conditions.

%%%%%%%%%%%%%%%%%%%%%%%%%%%%%%%%%%%%%%%%%%%%%%%%%%%%%%%%%%%%%%

\section*{Acknowledgements}

We would like to thank J.~Alwall, A.~Barr, C.~B.~Park and O.~Mattelaer
for useful discussions. This project was
supported in part by the National Science
Foundation under grant PHY-0854782.

%%%%%%%%%%%%%%%%%%%%%%%%%%%%%%%%%%%%%%%%%%%%%%%%%%%%%%%%%%%%%%

\section*{Appendix: Model cross sections}

The following table lists the tree-level parton-level production cross sections
$\sigma_{\rm prod}$ for the process $pp \to Y^+Y^- \to \ell^+\ell^- X^0
\bar{X}^0\, (\ell = e,\mu)$, for the 11 independent
combinations from Tab.~\ref{tab:models}. Also shown are  the measurable cross
sections $\sigma_{\rm meas}$ after inclusion of detector effects and the cuts in
eq.~\eqref{eq:cuts}. The cross sections have been computed with {\sc CompHEP}.

\renewcommand{\arraystretch}{1.2}
\begin{center}
\begin{tabular}{|l||c|c|}
\hline
Combination & $\sigma_{\rm prod}$ [fb] & $\sigma_{\rm meas}$ [fb] \\
\hline\hline
1 & \phantom{00}3.62 & \phantom{00}1.45 \\
\hline
2 & \phantom{00}8.50 & \phantom{00}3.36 \\
\hline
3 & \phantom{00}9.65 & \phantom{00}3.11 \\
\hline
4 & \phantom{0}41.4 & \phantom{0}11.45 \\
\hline
5 & \phantom{0}41.4 & \phantom{0}11.70 \\
\hline
6 & \phantom{0}41.4 & \phantom{0}14.05 \\
\hline
7 & \phantom{0}89.6 & \phantom{0}25.0 \\
\hline
8 & \phantom{0}29.9 & \phantom{00}8.47 \\
\hline
9 & \phantom{0}89.6 & \phantom{0}31.4 \\
\hline
10 & 112 & \phantom{0}31.2 \\
\hline
11 [$M_{\hat{Q}}$=0.5~TeV]& 179 & \phantom{0}48.3 \\
\hline
11 [$M_{\hat{Q}}$=1~TeV]& 445 & 137 \\
\hline
\end{tabular}
\end{center}

%%%%%%%%%%%%%%%%%%%%%%%%%%%%%%%%%%%%%%%%%%%%%%%%%%%%%%%%%%%%%%
%%%%%%%%%%%%%%%%%%%%%%%%%%%%%%%%%%%%%%%%%%%%%%%%%%%%%%%%%%%%%%

\end{document}